\documentclass[twoside,twocolumn,english,prl,superscriptaddress]{revtex4-1}
\usepackage[T1]{fontenc}
\usepackage[latin9]{inputenc}
\usepackage{bm}
\usepackage{amsmath}
\usepackage{graphicx}

\makeatletter
\usepackage[colorlinks,linkcolor=blue,anchorcolor=blue,citecolor=red]{hyperref}
\makeatother

\usepackage{babel}
\begin{document}
\title{Non-diagonal disorder enhanced topological properties of graphene
with laser irradiation}
\author{Tao Qin}
\email{taoqin@ahu.edu.cn}

\affiliation{School of Physics and Optoelectronics Engineering,
Anhui University, Hefei, Anhui Province 230601, People\textquoteright s
Republic of China}
\author{Pengfei Zhang}
\affiliation{School of Physics and Optoelectronics Engineering,
Anhui University, Hefei, Anhui Province 230601, People\textquoteright s
Republic of China}
\author{Guoao Yang}
\affiliation{School of Physics and Optoelectronics Engineering,
Anhui University, Hefei, Anhui Province 230601, People\textquoteright s
Republic of China}
\begin{abstract}
Laser irradiation, as a versatile tool to tune topological properties
of electronic systems, is under intensive studies. Experimentally,
laser irradiation induced anomalous Hall effect in graphene has been
observed (McIver et al., Nat. Phys. 16, 38 (2020)). Disorder is ubiquitous
in real materials, and it has been shown that diagonal disorders,
i.e., onsite disorder, can enhance topological properties of time-periodically
driven quantum materials (Titum et al., Phys. Rev. Lett. 114, 056801
(2015)). Here, we investigate circularly polarized laser irradiated
graphene with non-diagonal disorders, i.e., disordered tunneling,
and find that disorder can induce nontrivial topological properties,
characterized by Bott index and the real-space Chern number. Moreover,
we show that one can turn on the laser irradiation non-adiabatically
to drive the disordered graphene into non-trivial topological phase.
It is a scheme which is especially interesting for experimental implementations.
\end{abstract}
\maketitle

\section{Introduction}

Quantum materials with non-trivial topology have attracted great attention
in last decades. Laser irradiation, as an effective way of Floquet
engineering  to introduce time-periodic driving to a static system,
is a powerful way to tune the topological properties of solid state
systems. Aoki and Oka~\citep{oka_photovoltaic_2009} have pioneered
the prediction that a circularly polarized laser can open a gap in
the Dirac cone of graphene. This work sparked series of work in the
direction of tuning topological properties with polarized laser. Kitagawa
et al.~\citep{kitagawa_transport_2011} pointed out that it was,
in fact, a realization of the Haldane model~\citep{haldane_model_1988}
in the regime of nonequilibrium physics. Lindner el al.\citep{lindner_floquet_2011},
also proposed that using irradiation at microwave frequencies, a non-trivial
topological phased can be induced in a semiconductor quantum well.
Torres et al. carried out detailed work\citep{calvo_tuning_2011,dal_lago_floquet_2015,foa_torres_multiterminal_2014,perez-piskunow_floquet_2014,perez-piskunow_hierarchy_2015,torres_kubo_2005}
about transport properties of multiterminal graphene irradated by
laser, and proposed possible experimental schemes. Furthermore, in
the context of quantum simulation \citep{bukov_universal_2015,eckardt_colloquium:_2017,qin_charge_2018}
and quantum materials \citep{tokura_emergent_2017}, for both non-interacting
and interacting cases, lots of interesting proposals on Floquet systems
have been made. The theoretical insight is experimentally carried
out firstly in the platform of ultracold atoms in hexagonal optical
lattices~\citep{jotzu_experimental_2014,flaschner_observation_2017,qin_charge_2018}.
Remarkably, using an ultrafast circularly polarized laser pulse~\citep{mciver_light-induced_2020},
McIver et al. observed the anomalous Hall effect in graphene. Nevertheless,
the anomalous Hall conductance is not fully quantized.

Disorder, ubiquitous in real materials, plays an important role to
enhance topological properties. Topological Anderson insulator (TAI)
is predicted in Ref.\citep{li_topological_2009} and experimentally
realized using one-dimensional disordered ultracold atoms~\citep{meier_observation_2018}.
By introducing diagonal disorders, i.e., onsite disorder, to circularly
polarized laser irradiated graphene, Titum el al., \citep{titum_disorder-induced_2015}
identified a non-trivial topological phase induced by disorder, for
which the bulk state is totally localized \citep{titum_anomalous_2016},
dubbed as Floquet topological Anderson insulator (FTAI).

These progresses motivate us to ask whether non-diagonal disorder,
i.e., disordered tunneling, can enhance topological properties of
a circularly laser irradiated graphene, and more importantly, how
to implement the laser irradiation experimentally. Here, we theoretically
investigated a circularly polarized laser irradiated graphene in the
presence of non-diagonal disorders in the hopping terms. A non-trivial
topological phase induced by disorder has been identified. We characterized
its topological properties with Bott index \citep{loring_disordered_2010,loring_k-theory_2015}
and the real space Chern number \citep{bianco_mapping_2011,tran_topological_2015},
and demonstrated that it could be realized at a moderate frequency.
For the possible experimental implementation, we showed that the system
could be driven into a non-trivial topological phase by non-adiabatically
ramping up of the driving, in sharp contrast to the adiabatically
turning on of the driving for the pristine case \citep{dalessio_dynamical_2015}.
We stress that it is a scenario which is especially appealing experimentally.

The rest of the manuscript is organized as follows. In Sec. II, we
present the model and topological invariants we have investigated.
In Sec. III, results on ramping up of the laser irradiation are shown
and discussed. A conclusion and outlook is given in Sec. IV.

\section{The model and method}

In this section, we firstly present the model for the non-interacting
fermions in the circularly polarized laser irradiated graphene as,
\begin{align}
H & =-J\sum_{\left\langle i\alpha,j\beta\right\rangle }\left(1+W\eta_{ij}\right)e^{iA_{ij}}c_{i\alpha}^{\dagger}c_{j\beta}+M\sum_{i,\alpha}\sigma_{\alpha\alpha}^{z}c_{i\alpha}^{\dagger}c_{i\alpha}\label{eq:floquet_honey}
\end{align}
where $J$ is the hopping amplitude, $A_{ij}=\bm{A}\left(t\right)\cdot\left(\bm{r}_{i}-\bm{r}_{j}\right)$
for the bond between site $i$ and $j$, and $\bm{A}=A_{0}\left(\cos\left(\Omega t\right),\sin\left(\Omega t\right)\right)$.
$\Omega$ is the driving frequency. $c_{i\alpha}$ ($c_{i\alpha}^{\dagger}$)
is the fermion annihilation (creation) operator at atom $\alpha$
in unit cell $i$. $\alpha,\beta=A,B$ label sub-lattices. $M$ is
the staggered potential. Effectively, in the high-frequency limit
the Hamiltonian~(\ref{eq:floquet_honey}) without disorder is the
Haldane model~\citep{jotzu_experimental_2014,flaschner_observation_2017}.
Non-zero disorder strength $W$ introduces non-diagonal disorder,
respectively. $\eta_{ij}$ is the random number uniformly distributed
in the range $[-0.5,0.5]$.

We obtain quasi-energy spectrum of the Floquet Hamiltonian~(\ref{eq:floquet_honey})
by~\citep{kitagawa_topological_2010,titum_disorder-induced_2017},
\begin{equation}
H_{\mathrm{eff}}=\frac{i}{T}\log U\left(T,0\right),\label{eq:Heff}
\end{equation}
where $U\left(T,0\right)=e^{-i\int_{0}^{T}H\left(t\right)dt}=\prod_{n=0}^{N-1}U\left(t_{n}+\Delta T,t_{n}\right)$,
with $U\left(t_{n}+\Delta T,t_{n}\right)=e^{-iH\left(t_{n}\right)\Delta T}$,
$t_{n}=n\Delta T$, $\Delta T=\frac{T}{N}$ and $T=\frac{2\pi}{\Omega}$.
It is different from the usual method for the effective Hamiltonian~\citep{bukov_universal_2015},
and one can obtain the exact quasi-energy spectrum in $\left[-\frac{\Omega}{2},\frac{\Omega}{2}\right]$.
We have a remark on the time-evolution operator $U\left(t_{n}+\Delta T,t_{n}\right)=e^{-iH\left(t_{n}\right)\Delta T}$,
which can be numerically implemented in a more efficiently way by
introducing the Trotter--Suzuki decomposition \citep{dalessio_dynamical_2015}
as
\begin{align*}
U\left(t_{n}+\Delta T,t_{n}\right) & =e^{-iH_{1}\left(t_{n}+\frac{\Delta T}{2}\right)\frac{\Delta T}{2}}e^{-iH_{0}\Delta T}\\
 & \cdot e^{-iH_{1}\left(t_{n}+\frac{\Delta T}{2}\right)\frac{\Delta T}{2}},
\end{align*}
 where $H_{0}=M\sum_{i,\alpha}\sigma_{\alpha\alpha}^{z}c_{i\alpha}^{\dagger}c_{i\alpha}$
and $H_{1}=-J\sum_{\left\langle i\alpha,j\beta\right\rangle }\left(1+Wr_{ij}\right)e^{iA_{ij}}c_{i\alpha}^{\dagger}c_{j\beta}$.

We focus on topological properties of the bands between $-\frac{\Omega}{2}$
and 0. Alternatively, one can obtain the quasi-energy spectrum by
transforming the Hamiltonian~(\ref{eq:floquet_honey}) into the Floquet
space~\citep{rudner_anomalous_2013,eckardt_high-frequency_2015}
by $\mathcal{H}^{mm^{\prime}}=m\Omega\delta_{\alpha\alpha^{\prime}}\delta_{mm^{\prime}}+\frac{1}{T}\int_{0}^{T}dte^{-i\left(m-m^{\prime}\right)\Omega t}H\left(t\right)$,
where $m$ and $n$ are Floquet indices, and keeping the quasi-energy
spectrum between $-\frac{\Omega}{2}$ and 0. In this work, we mainly
use the Hamiltonian defined in Eq.~(\ref{eq:Heff}), and the Hamiltonian
$\mathcal{H}^{mm^{\prime}}$ in the Floquet space serves as a cross-check.
For a non-interacting disordered system, we use the Bott index $\bar{C}_{b}\left(-\frac{\Omega}{2},0\right)$
as a function of disorder strength to characterize different topological
phases in the real space. In a time-independent Hamiltonian, it is
shown that the Bott index is equivalent to the Chern number~\citep{toniolo_equivalence_2017}.
For a specific disorder configuration, the Bott index is defined by~\citep{loring_disordered_2010,loring_k-theory_2015},
\begin{equation}
C_{b}\left(-\frac{\Omega}{2},0\right)=\frac{1}{2\pi}\mathrm{Im}\left[\mathrm{Tr}\left\{ \mathrm{ln}\left(\tilde{U}_{Y}\tilde{U}_{X}\tilde{U}_{Y}^{\dagger}\tilde{U}_{X}^{\dagger}\right)\right\} \right]
\end{equation}
where $\tilde{U}_{X\left(Y\right)}=PU_{X\left(Y\right)}P$. The unitary
matrices $U_{X}=\exp\left(i2\pi X/L_{x}\right)$ and $U_{Y}=\exp\left(i2\pi Y/L_{y}\right)$,
where $X$($Y$) is diagonal matrix of the $x$ ($y$) coordinate
of all the lattice sites. $P$ is the projector operator to project
the quasi-energy spectrum into the range between $-\frac{\Omega}{2}$
and 0.

A second quantity can be used to characterize the topological properties
of the disordered lattice system is the Chern number in the real space~\citep{bianco_mapping_2011,tran_topological_2015},
which is perfectly suited for a disordered case. Here we adopt in
our calculations the formula for the real space Chern number presented
in Ref.\citep{tran_topological_2015}.

\section{Results and discussions}

\subsection{The Bott index}

\begin{figure}[h]
\includegraphics[scale=0.55]{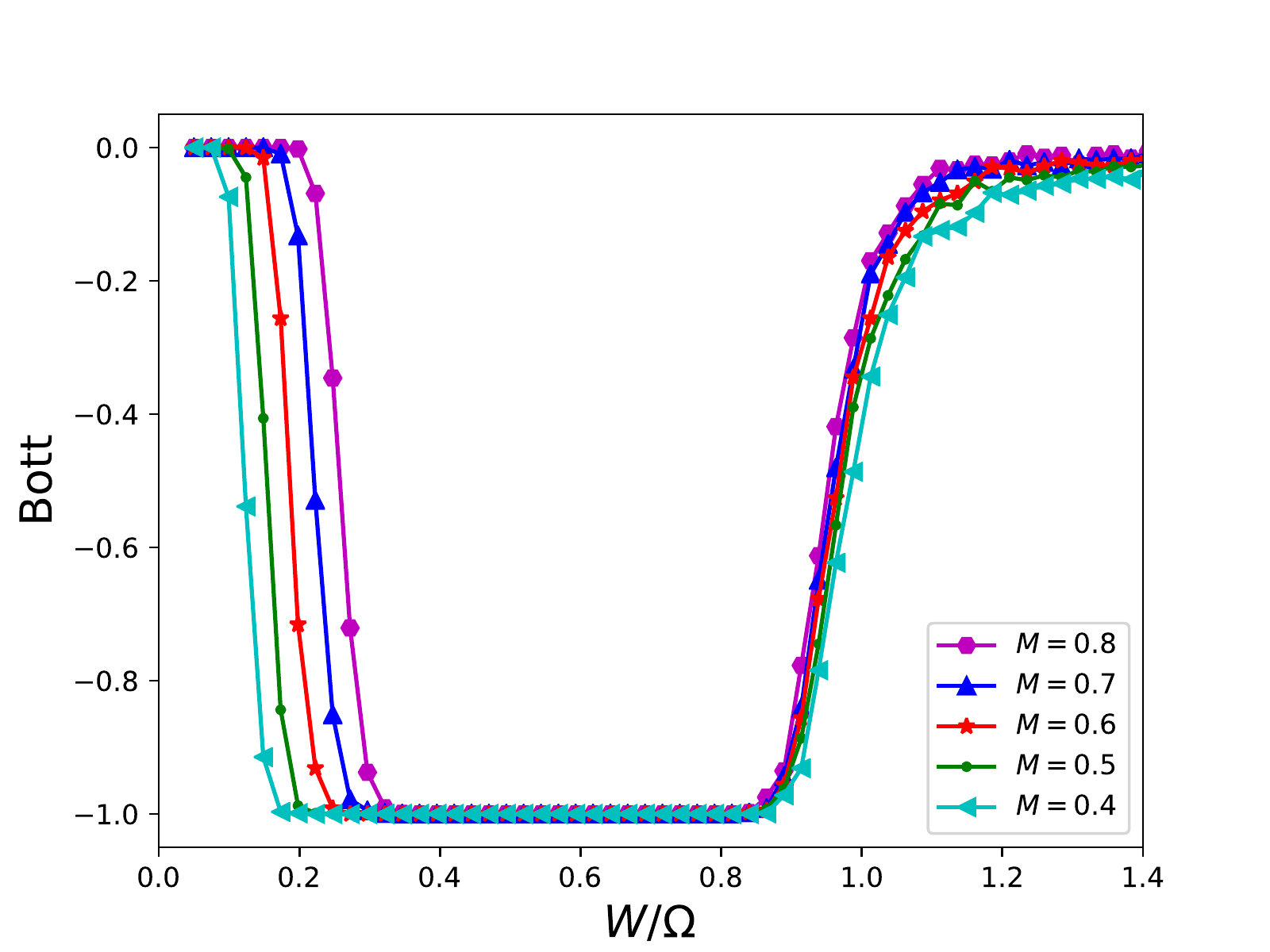}

\caption{\label{fig:Bott-W-m} The Bott index $\bar{C}_{b}\left(-\frac{\Omega}{2},0\right)$
for the Hamiltonian~(\ref{eq:Heff}) as a function of disordered
tunneling strength for a lattice of 24$\times$24 with different staggered
potential $M$. The driving frequency $\Omega=6$. $A_{0}=1.6$. A
periodical boundary condition is adopted. In this and the following
calculations, the energy unit is $J\mathcal{J}_{0}\left(A_{0}\right)$
with $\mathcal{J}_{0}\left(A_{0}\right)$ as the zeroth order Bessel
function of the first kind. 960 realizations are calculated.}
\end{figure}

To characterize topological properties of laser-irradiated graphene,
we firstly present the results for the Bott indices as a function
of the disorder strength. In this work we choose the energy unit as
$J\mathcal{J}_{0}\left(A_{0}\right)$ with $\mathcal{J}_{0}\left(A_{0}\right)$
as the zeroth order Bessel function of the first kind. In Fig.~\ref{fig:Bott-W-m},
we show that the disordered tunneling can induce non-trivial topological
phase for the driving frequency $\Omega=6$. The Bott index $-1$
shows the difference in the number of edge states in the gap $-\frac{\Omega}{2}$
and $0$. This phenomena exists for a relatively large range of staggered
potentials. For different staggered potentials, the critical $M$
is different. The introduction of the disordered tunneling effectively
changes the staggered potential. For a certain staggered potential
$M=0.8$, we explore the properties of the Bott index for different
driving frequencies. The phenomena that the disordered tunneling can
induce non-trivial topological phase disappears for a relatively high
driving frequency. We do not go to a very low driving frequency, which
may be corresponding to the gap closing at the energy $-\frac{\Omega}{2}$
and 0. We would like to point out that the Bott index as $-1$ or
$1$ is not essential, as can be tuned by $A_{ij}$ in the Hamiltonian
(\ref{eq:floquet_honey}).

\begin{figure}[h]
\includegraphics[scale=0.55]{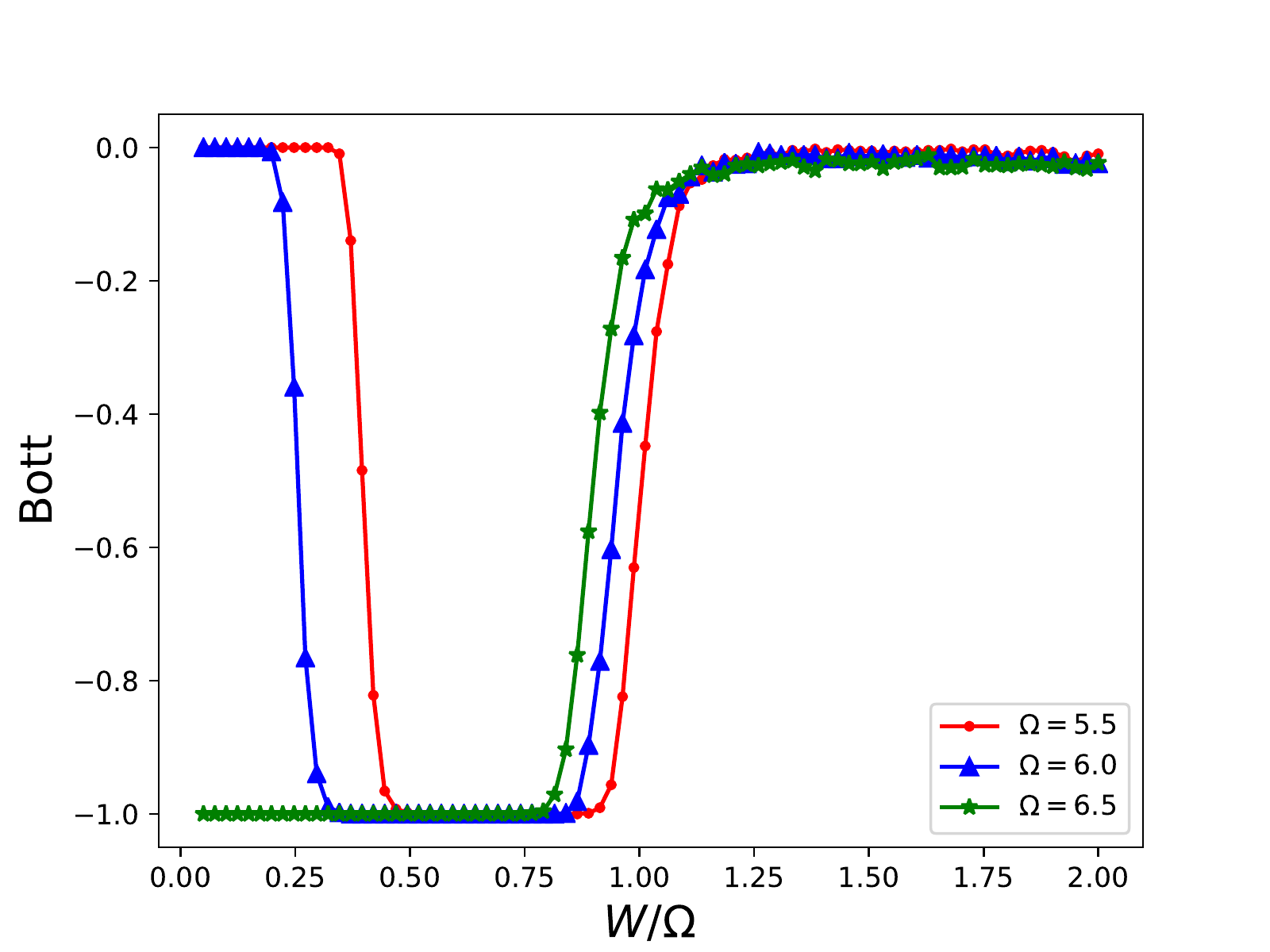}

\caption{\label{fig:Bott-W-Ome} The Bott index $\bar{C}_{b}\left(-\frac{\Omega}{2},0\right)$
for the Hamiltonian~(\ref{eq:Heff}) with disordered tunneling for
a lattice of 24$\times$24 for different driving frequencies. A periodical
boundary condition is adopted. The staggered potential $M=0.8$. $A_{0}=1.6$.
960 realizations of disorder are calculated.}
\end{figure}

\subsection{The real-space Chern number}

\begin{figure}
\includegraphics[scale=0.55]{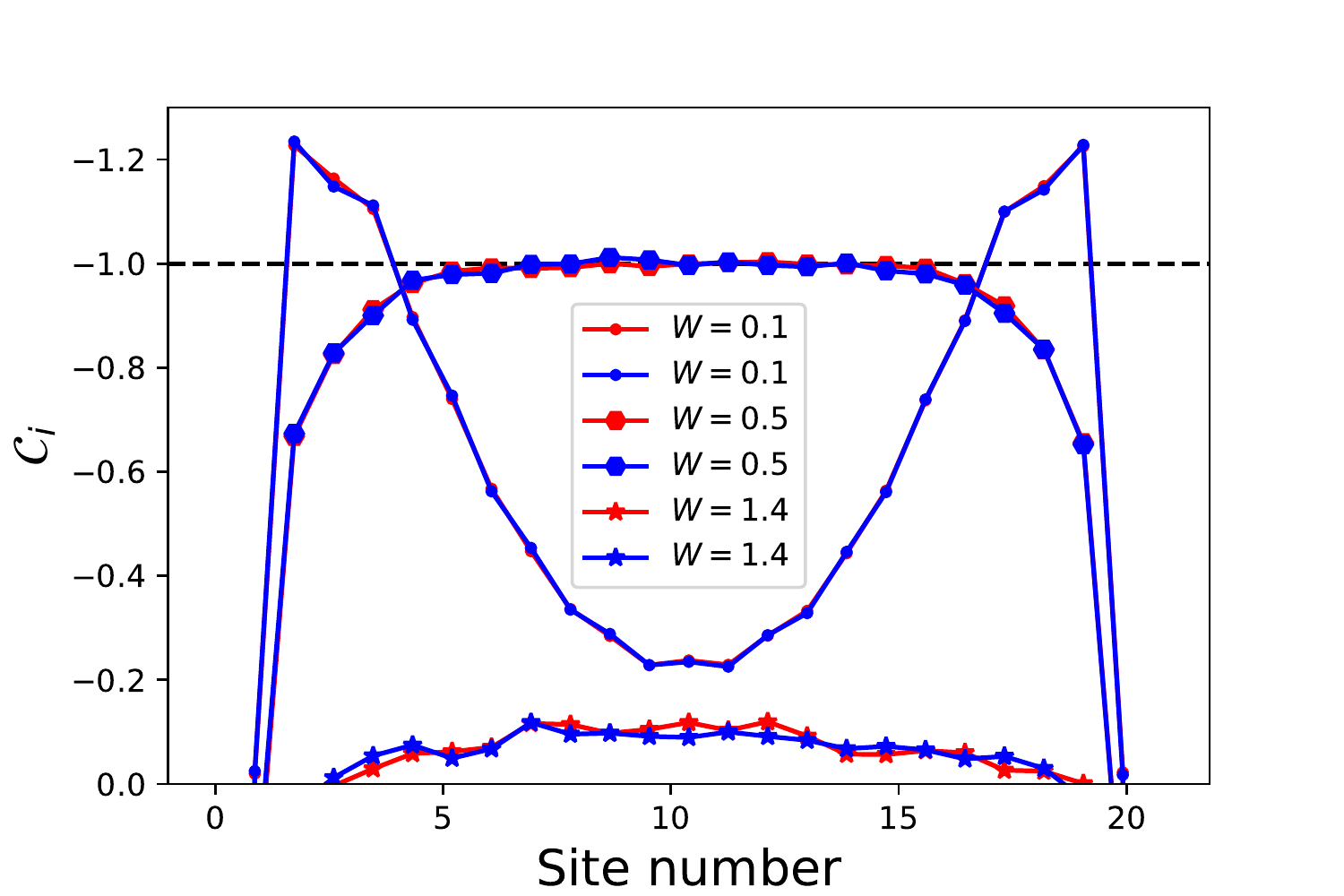}

\caption{\label{fig:Chern-marker}The Chern marker $\mathcal{C}_{i}$ for a
lattice of 25$\times$24 for different non-diagonal disorder strengths.
The blue and red lines are Chern makers for 12-th and 13-th rows of
lattice sites. An open boundary condition is adopted. The driving
frequency $\Omega=6$. $A_{0}=1.6$. 960 realizations of disorder
are calculated.}
\end{figure}

To further confirm the non-trivial topological properties due to the
non-diagonal disorder, we calculate the real-space Chern number, which
is dubbed as the Chern marker. For a hexagonal lattice of 25$\times$24
sites with open boundary conditions, in Fig.~\ref{fig:Chern-marker},
we show Chern markers versus the column indices for sites of the 12-th
and 13-th rows in the bulk. For bulk sites in the 12-th and 13-th
rows, the Chern markers are nearly uniform, meaning that they are
well-defined topological index. More importantly, consistent with
results for the Bott indices, in the non-trivial topological phase
induced by non-diagonal disorder ($W=0.5$ in Fig.~\ref{fig:Chern-marker})
Chern markers are about 1, in sharp contrast to the phase with trivial
topology where it is largely deviated from 1.

\subsection{Ramping-up of the time-periodical driving and possible experimental
realization}

\begin{figure}
\includegraphics[scale=0.6]{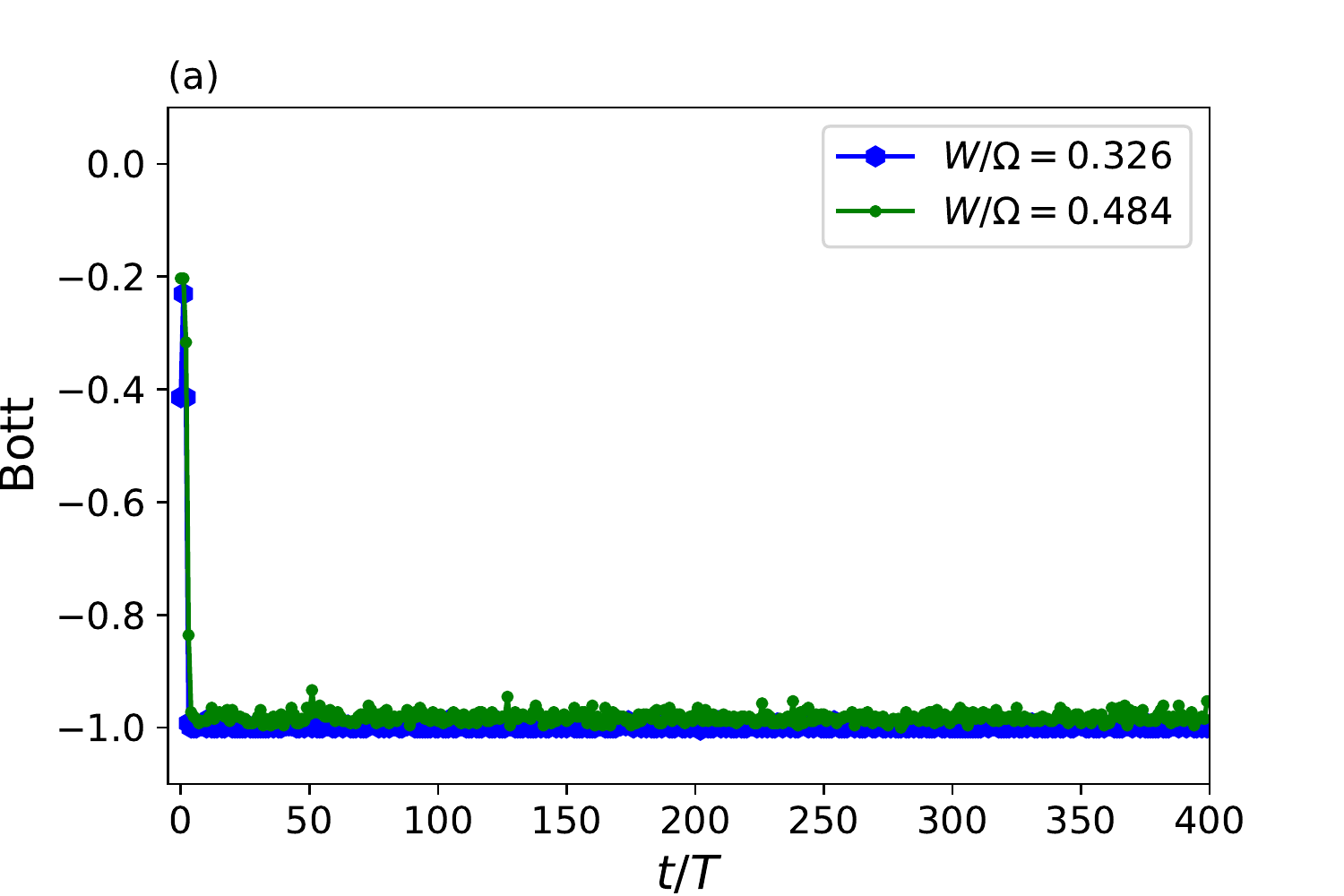}

\includegraphics[scale=0.6]{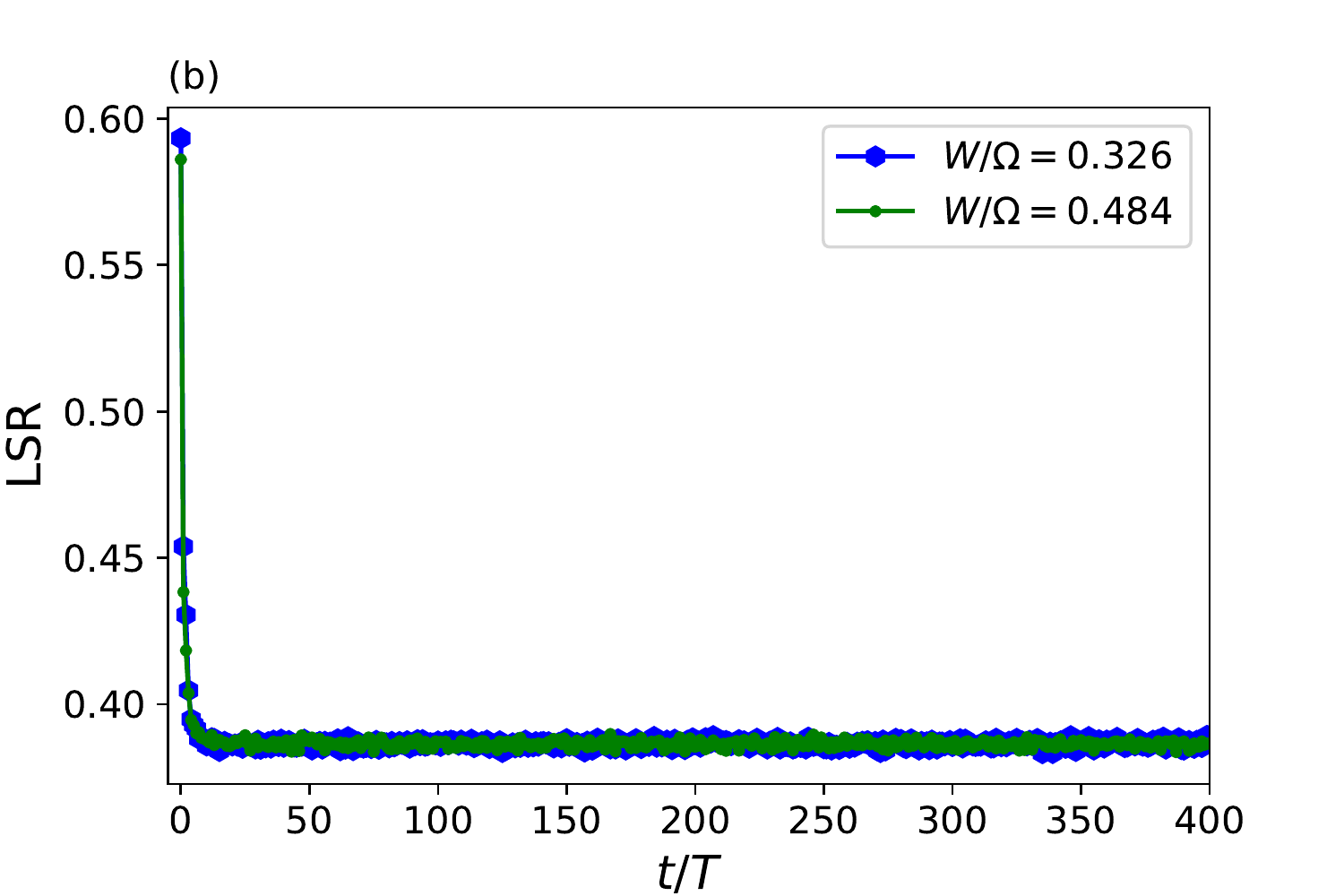}

\caption{\label{fig:time} (a) With a quenched ramp up of the laser irradiation,
the time evolution of the Bott index $\bar{C}_{b}\left(-\frac{\Omega}{2},0\right)$
and level spacing ratio (LSR) for the Hamiltonian~(\ref{eq:Heff})
with disordered tunneling from an initial state with trivial topologyfor
a lattice of 18$\times$18 for different disorder strengths. A periodical
boundary condition is adopted. The staggered potential $M=0.7$. $A_{0}=1.6$.
$N_{0}=400$, $N=150$. 256 realizations of disorder are calculated.}

\end{figure}

With disordered tunneling present, we investigate the possible ways
of ramping up for the laser irradiation. In general, the driving from
lasers can be turned on adiabatically or non-adiabatically. For a
pristine system, it is the general wisdom that one needs to turn on
the laser irradiation adiabatically \citep{dalessio_dynamical_2015}
to drive the system into non-trivial topological phase. For disordered
system, we investigate the time evolution from an initial state of
a disordered static system with trivial topological properties as
follows,
\begin{equation}
\left|\psi\left(t\right)\right\rangle =\prod_{n_{2}=0}^{N_{0}}\prod_{n_{1}=0}^{N}U\left(t_{n}+\Delta T,t_{n}\right)\left|\psi\left(t=0\right)\right\rangle
\end{equation}
with $N_{0}$ the number total periods investigated and $N$ the number
of a single period divided. $t_{n}=n_{2}T+n_{1}\Delta T$. As we have
remarked before, we calculate the time evolution operator $U\left(t_{n}+\Delta T,t_{n}\right)$
with the Trotter-Suzuki decomposition. We obtain the Bott index and
level spacing ratio (LSR) \citep{oganesyan_localization_2007} at
$t=nT$ in a stroboscopic way. The key result in this section is shown
in Fig.\ref{fig:time}(a) that one needs to turn on the laser irradiation
in a \textit{non-adiabatical} way to enhance the topological properties,
in stark contrast to a totally pristine driven system, where one needs
to turn on the driving adiabatically\citep{dalessio_dynamical_2015}.
Driving by laser irradiation tends to heat the system up. If it is
turned on adiabatically, the system is not heated up before it enters
into topological phase. We attribute it to the phenomena of coherent
destruction\citep{grossmann_coherent_1991} that driving contributes
to localize the system. Once that the system enters into the phase
with Bott index as $-1$, the \textit{bulk} state is totally localized,
with LSR as about 0.39 in Fig.\ref{fig:time}(b) \citep{titum_anomalous_2016,atas_distribution_2013},
a necessary condition for FTAI. Therefore, disorder helps the system
to become topologically non-trivial by localizing bulk states. The
final resolution for the system is an interplay among heating, localization
and topology.

\begin{figure}
\includegraphics[scale=0.6]{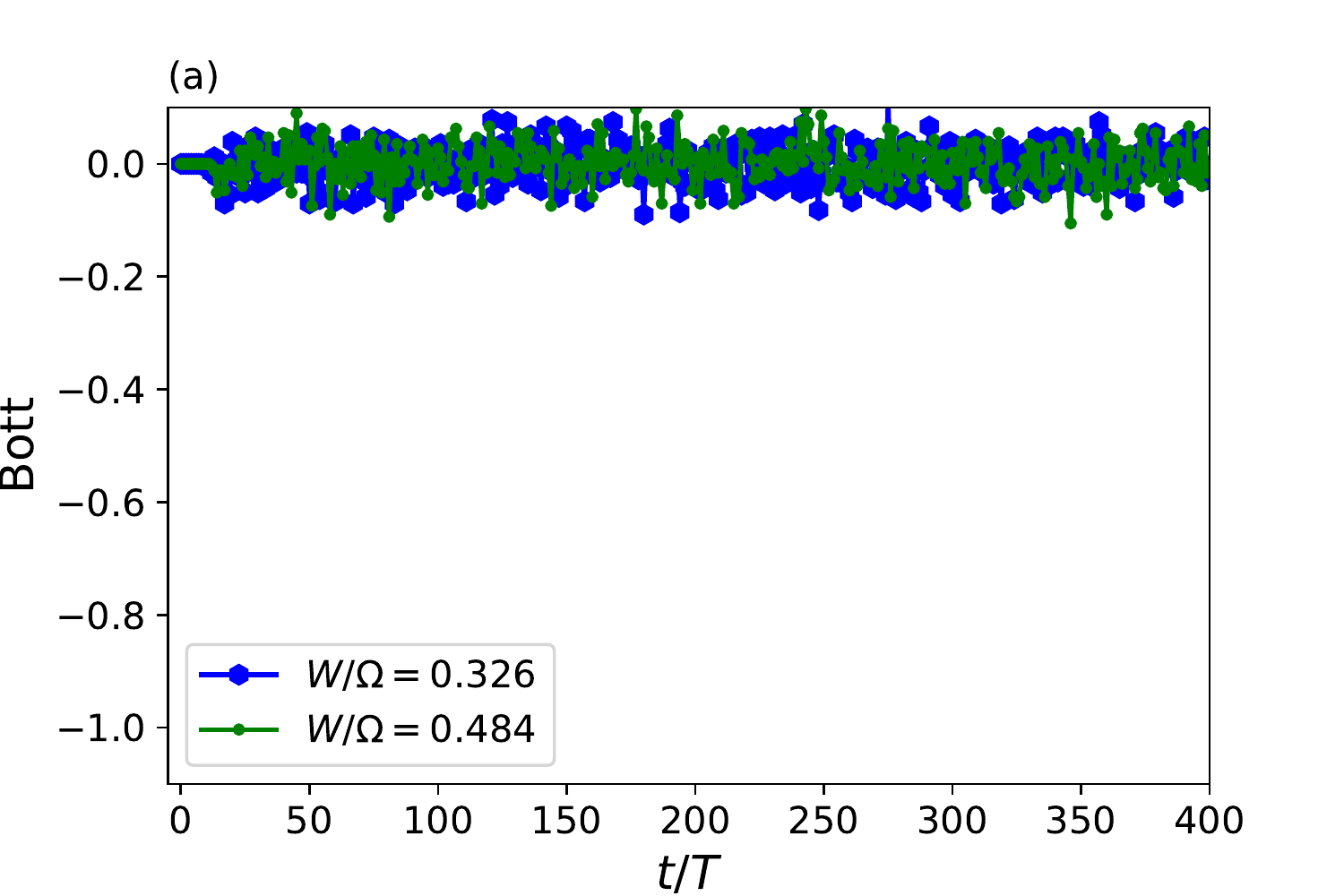}

\includegraphics[scale=0.6]{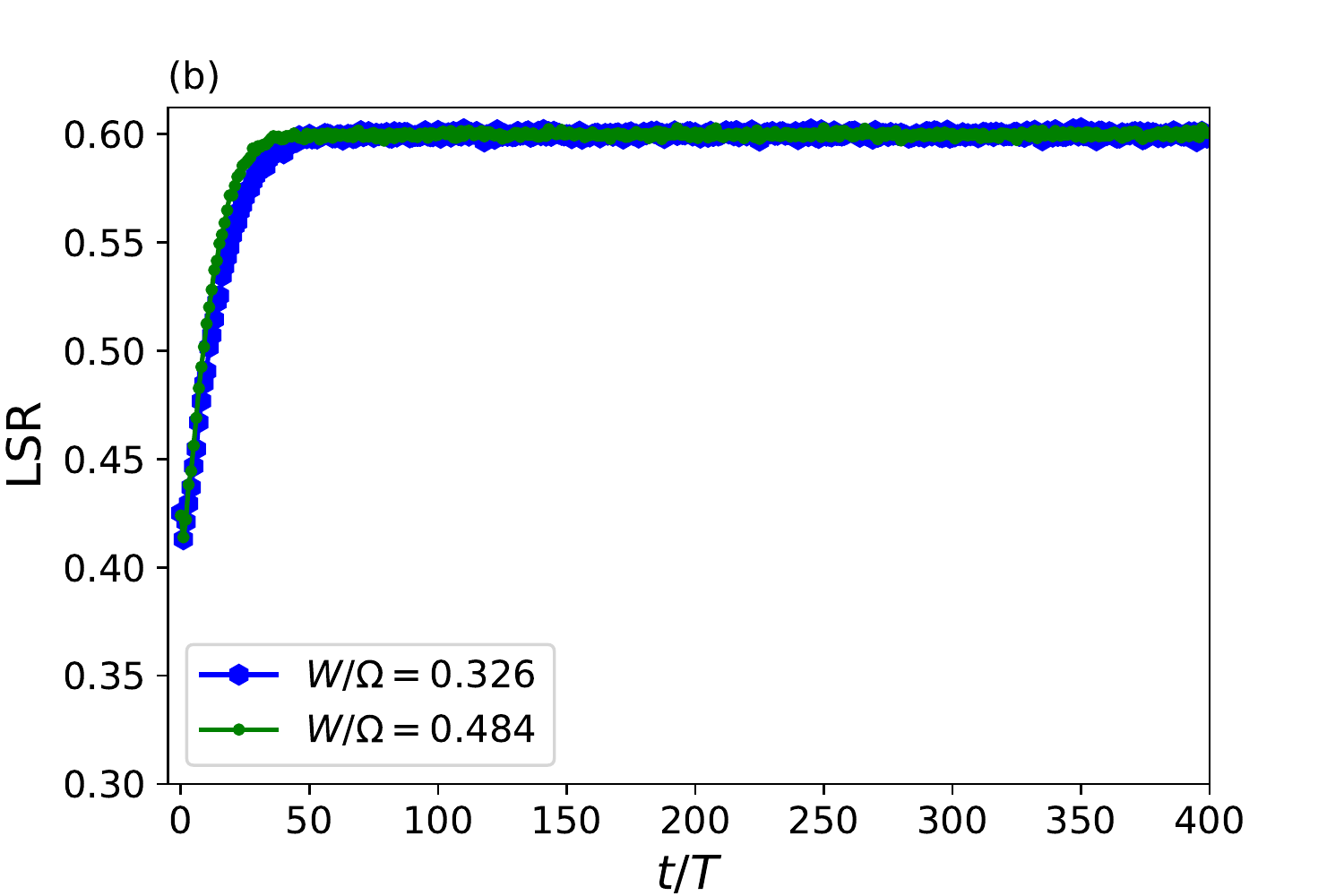}

\caption{\label{fig:linear_time} (a) With a linear ramp up of the laser irradiation
by setting $\bm{A}=A_{0}\frac{t}{\tau}\left(\left(\cos\left(\Omega t\right),\sin\left(\Omega t\right)\right)\right)$,
the time evolution of the Bott index $\bar{C}_{b}\left(-\frac{\Omega}{2},0\right)$
and level spacing ratio (LSR) for the Hamiltonian~(\ref{eq:Heff})
with disordered tunneling from an initial state with trivial topology
for a lattice of 18$\times$18 for different disorder strengths. $\tau=200T$.
The other parameters are the same as these in Fig.\ref{fig:time}. }
\end{figure}

To further clarify the issue of heating, localization and topology,
we present in Figs.\ref{fig:linear_time} the same physical quantities
but with a linear ramp-up of the laser irradiation by setting $\bm{A}=A_{0}\frac{t}{\tau}\left(\left(\cos\left(\Omega t\right),\sin\left(\Omega t\right)\right)\right)$
with $\tau=200T$. It is clear in Figs.\ref{fig:linear_time} that
for different disorder strengths, the system ends up with Bott indices
as 0 and LSR as about 0.6, meaning that the system is topologically
trivial and delocalized\citep{titum_anomalous_2016,atas_distribution_2013}.
It shows that the system can absorb energy from the driving effectively
by gradually turning on the laser irradiation, leading to a fully
delocalized phase. Therefore, localization is necessary for the system
to enter into the non-trivial topological phase, and otherwise the
system would be heated up.

The important experimental progress \citep{mciver_light-induced_2020}
shows the power of laser irradiation. However, we would like to point
out that the anomalous Hall conductance is not fully quantized. One
possible way to enhance it is to introduce disorders, whether non-diagonal
disorder in this work or diagonal disorder explored in literatures
\citep{titum_disorder-induced_2015,titum_anomalous_2016}, for both
disorders are ubiquitous in a real materials. A second way to realize
the Hamiltonian with non-diagonal disorder is to use electric circuits
\citep{nakata_circuit_2012}.

\section{Conclusion and Outlook}

We have investigated a laser irradiated graphene system with disordered
tunneling. By characterizing the topological properties with the Bott
index and the local Chern marker, we find that topological properties
of the system is enhanced by disordered tunneling. Moreover, we have
shown an appealing experimental scenario that one needs to turn on
the irradiation in a quenched way to drive the system into the non-trivial
topological phase.
\begin{acknowledgments}
We thank Prof. Dr. Walter Hofstetter in Frankfurt University in Germany,
Dr. Feng Mei at Shanxi University in China, and Dr. Yong Xu at Tsinghua
Univeristy in China for insightful discussions and communications.
This work is supported by the National Natural Science Foundation
of China (U2032164), and the start-up fund from Anhui University in
China.
\end{acknowledgments}

\bibliographystyle{apsrev4-1}
%\bibliography{My_Library}
%merlin.mbs apsrev4-1.bst 2010-07-25 4.21a (PWD, AO, DPC) hacked
%Control: key (0)
%Control: author (72) initials jnrlst
%Control: editor formatted (1) identically to author
%Control: production of article title (-1) disabled
%Control: page (0) single
%Control: year (1) truncated
%Control: production of eprint (0) enabled
%

\end{document}